 \documentclass[final,5p,times,twocolumn]{elsarticle}

\usepackage{graphicx}

\usepackage{amssymb}
\usepackage{amsthm}

\newcounter{bla}

\journal{Computer Physics Communications}

\usepackage{amsmath, latexsym}
\usepackage{amsfonts}
\usepackage{physicsShortCut}
\usepackage{tdgcmNotations}
\usepackage[colorlinks=true]{hyperref}
\usepackage{url}

\graphicspath{{figures/}}

\begin{document}

\begin{frontmatter}



\title{\theCode-1.0: A finite element solver for the time dependent generator coordinate method with the Gaussian overlap approximation}


\author[a,b]{D. Regnier}
\author[b]{M. Verri\`ere}
\author[b]{N. Dubray}
\author[a]{N. Schunck\corref{author}}

\cortext[author] {Corresponding author.\\\textit{E-mail address:} schunck1@llnl.gov}
\address[a]{Nuclear and Chemical Science Division, Lawrence Livermore National Laboratory, Livermore, CA 94551, USA}
\address[b]{CEA, DAM, DIF, 91297 Arpajon, France}

\begin{abstract}
We describe the software package \theCode \ that solves the equations of the time-dependent generator coordinate method (TDGCM) in $N$-dimensions ($N \geq 1$) under the Gaussian overlap approximation.
The numerical resolution is based on the Galerkin finite element discretization of the collective space and the Crank-Nicolson scheme for time integration.
The TDGCM solver is implemented entirely in C++. Several additional tools written in C++, Python or bash scripting language are also included for convenience.
In this paper, the solver is tested with a series of benchmarks calculations.
We  also demonstrate the ability of our code to  handle a realistic calculation of fission dynamics.
\end{abstract}

\begin{keyword}
\theCode; Finite element method; Generator coordinate method; Gaussian overlap approximation; Nuclear fission; 
\end{keyword}

\end{frontmatter}



{\bf PROGRAM SUMMARY/NEW VERSION PROGRAM SUMMARY}

\begin{small}
\noindent
{\em Program Title:}     \theCode-1.0                                     \\
 {\em Journal Reference:}                                      \\
{\em Catalogue identifier:}                                   \\
{\em Licensing provisions:}                                   \\
{\em Programming language: } C++                                  \\
{\em Computer:} Intel Xeon, Intel Core                                              \\
{\em Operating system: } LINUX                                      \\
{\em RAM:}
Memory usage depends on the number of nodes in the calculation mesh as well as on the degree of the interpolation polynomials.
For a 1D calculation with linear polynomials on a mesh with 600 nodes, memory usage is approximately 3.3 MB; in a realistic simulation of fission on a 2D mesh with quadratic polynomials and $1.3 \, 10^5$ nodes, it reaches 1.5 GiB.
\\
{\em Number of processors used:}                              \\
  The code is multi-threaded based on the OpenMP API specification for parallel programming.
  Any number of threads may be specified by the user.
  \\
{\em Keywords:} \theCode; Finite element method; Generator coordinate method; Gaussian overlap approximation; Nuclear fission;   \\
{\em Classification:}    17.23 Fission and Fusion Processes                                       \\
{\em External routines/libraries:} The solver itself requires the BLAS and LAPACK libraries, and a Fortran compiler with 
OpenMP support. Building the documentation requires DoxyGen-1.8.6 or higher. Building the  full set of tools also requires GSL, PETSc, SLEPc and 
Boost. In particular, environment variables PETSC\_DIR, PETSC\_ARCH, SLEPC\_DIR and SLEPC\_ARCH must be set. 
\\
{\em Nature of problem:}\\
Nuclear fission is a relatively slow process compared to the typical timescale of the intrinsic motion of the nucleons. In the adiabatic approximation, it can be described as a large amplitude collective motion driven by only a few collective degrees of freedom. In the time-dependent generator coordinate method (TDGCM), the nuclear wave-function is thus described as a time-dependent, linear superposition of basis functions in this collective space. Further assuming a Gaussian overlap approximation (GOA) for the basis functions, the time-dependent Schr\"{o}dinger equation can be reduced into a local, time-dependent, Schr\"{o}dinger-like equation in collective space. This is the TDGCM+GOA equation. Scission configurations are defined as a hyper-surface in the $N$-dimensional collective space. Fission fragment distributions are then computed by integrating over time the flux of the collective wave-packet across the scission hyper-surface. This microscopic approach to fission fragment distributions is fully quantum-mechanical.
   \\
{\em Solution method:}\\
\theCode \ solves the TDGCM+GOA equation by using the Galerkin finite element method to discretize the $N$-dimensional collective space, and the Crank-Nicolson scheme to solve for the time evolution. At each time step, this procedure requires solving a linear system of equation involving sparse, complex, symmetric matrices. \theCode\ employs an iterative QMR algorithm to perform matrix inversion.
   \\
{\em Restrictions:}\\
Although the program can operate in an arbitrary number of dimensions $N$, it has only been tested in practice on 1, 2 and 3 dimensional meshes.
   \\
{\em Unusual features:}\\
   \\
{\em Additional comments:}\\
The code has checkpointing capabilities: the collective wave-function, norm a
and energy kernels are stored on disk every $n$ iterations, ensuring that the 
program can resume where it stops.
   \\
{\em Running time:}\\
Running time grows linearly with the number of time-steps requested by the user. It is also highly dependent on the number of nodes in the space mesh. Two periods of a 1D harmonic oscillator (600 nodes, 800 time steps) are typically computed in a few seconds on one thread of a Intel(R) Core(TM) i5 CPU. A 2-dimensional realistic case of fission ($10^5$ nodes, $10^5$ time steps) requires roughly 10 hours on 10 threads of an Intel Xeon EP X5660 processor.
   \\
\end{small}

\section{Introduction}
\label{sec:introduction}

Induced nuclear fission plays an essential role in important societal applications ranging from stockpile science to critical assemblies for new generation nuclear reactors \cite{rising2013}. It is also one of the leading mechanisms determining the stability of super-heavy elements and the end point of nucleosynthesis \cite{grawe2007}. Many of these applications require the detailed knowledge of fission product yields (FPY), which may include the charge, mass, kinetic energy and excitation energy distribution of the fission fragments. In spite of recent technological advances, FPY measurements are not always possible, especially in very short-lived neutron-rich or heavy nuclei. Predictions based on theoretical models of fission are thus unavoidable. 

While there exists a number of powerful phenomenological or semi-microscopic models on the market, the consensus is that a truly predictive theory of fission should ultimately be based only on our knowledge of nuclear forces and quantum many-body methods. In this context, most of the effort in the last decades has been focused on describing the dynamics of induced fission in the time-dependent generator coordinate method (TDGCM) associated with the Gaussian overlap approximation (GOA)~\cite{griffin_collective_1957,reinhard_generator-coordinate_1987,ring_nuclear_2004}.
Under these assumptions, the original time-dependent many-body Schr\"odinger equation is reduced to a local Schr\"odinger-like equation that depends on only a few relevant collective variables. This approach was able to predict the characteristic times of low-energy induced fission~\cite{berger_microscopic_1984,berger_time-dependent_1991}. More recently, it was also successfully used to provide the first estimate of the mass and kinetic energy distributions of fission fragments for neutron-induced fission in the actinides region~\cite{goutte_microscopic_2005,younes_fragment_2012,younes_collective_2012}. 

Until now, the aforementioned calculations have been based on the discretization of the TDGCM equations using finite differences in a regularly meshed hyper-cube. Given the computational resources required by this simple scheme, fission dynamics has only been studied in 2-dimensional collective spaces. Yet, it is well-known from both semi-phenomenological and fully microscopic approaches that at least four or five collective variables play a role in the dynamics of fission~\cite{moeller_micmac_2001,younes_microscopic_2009,dubray_numerical_2012,schunck_microscopic_2014}. Although possible in theory, extending the current scheme to $N>2$ collective spaces would be prohibitive computationally. Similarly, increasing the fidelity of the calculation from $n_{0}$ points to $n$ points for all $N$ dimensions scales approximately like $(n/n_{0})^{N}$. A lot of this increase in computational cost would be  wasted in regions of the collective space far away from the scission configurations, where the resolution does not need to be very high. For these reasons, it is highly desirable to move to a more flexible, more scalable discretization scheme.

In this work, we thus introduce the code \theCode, which implements the Galerkin finite element method for the discretization of the TDGCM equation. This well-known method allows the use of irregularly-spaced meshes in the collective space, which results in turn in calculations scaling much more efficiently with the number of collective variables. In addition, standard p-refinement techniques, i.e., the use of higher-degree polynomial bases in each finite element, give a better control on the numerical precision of the calculations. Finally, there are virtually no restrictions in the number of collective variables used. The code \theCode \ has been tested up to $N=3$.

After a brief introduction to the TDGCM+GOA equations in section~\ref{sec:equationCollectiveDynamics}, 
we present in section~\ref{sec:numericalResolution} the numerical implementation in the code \theCode. The implementation is validated through a series of simple benchmarks discussed in section~\ref{sec:benchmarks}. The convergence of a realistic calculation of the fission of $^{240}$Pu is demonstrated in section~\ref{sec:fissionPu240}.
Finally, sections~\ref{sec:programmTheCode} and ~\ref{sec:inputsOutputs} give practical information on how to install and use the code \theCode.

\section{Fission dynamics in the TDGCM+GOA approach}
\label{sec:equationCollectiveDynamics}

In this section, we briefly recall how to obtain a collective, Schr\"odinger-like equation to describe low-energy nuclear dynamics. We also explain how fission fragment distributions can be extracted from the integration of the collective flux across the hyper-surface defining the scission configurations.

%
%

\subsection{The TDGCM+GOA equation}

We recall that the time evolution of a many-body quantum system is given by the time-dependent Schr\"{o}dinger equation, which is obtained from the variation $\delta\mathcal{S}[\Psi] = 0$ of the quantum mechanical action given by \cite{marquez2006}
\begin{equation}
\label{eq:shrodingerVariational}
\mathcal{S}[\Psi] = \int_{t_{0}}^{t_{1}}\frac{\langle \Psi(t) \,|\, \left[ \hat{H} -\hbar \frac{\partial}{\partial t}\right] \,|\, \Psi(t) \rangle}{\langle \Psi(t) | \Psi(t)\rangle},
\end{equation}
where $|\Psi(t)\rangle$ is the full many-body wave-function for the system. In most nuclear physics applications, the nuclear Hamiltonian $\hat{H}$ contains an effective two-body potential such as, e.g., the Skyrme or Gogny interaction. In the time-dependent generator coordinate method (TDGCM), the nuclear many-body wave function $|\Psi(t)\rangle$ takes the form\cite{griffin_collective_1957,reinhard_generator-coordinate_1987,ring_nuclear_2004}
\begin{equation}
\label{eq:gcmApprox}
|\Psi(t)\rangle = \int d\qVec\; f(\qVec, t)|\Psi(\qVec) \rangle.
\end{equation}
The functions $|\Psi(\qVec)\rangle$ are known many-body states parametrized by a vector of collective variables $\qVec$. In the context of fission, the $|\Psi(\qVec)\rangle$ are chosen as the solutions to the static Hartree-Fock-Bogoliubov (HFB) equations under a set of constraints $\qVec$. These constraints, which are the collective variables driving the fission process, can be expectation values of multipole moments, quantities related to pairing such as particle number fluctuations $\Delta N^{2}$, etc. Recall that the HFB solutions at point $\qVec$ are entirely characterized by the one-body density matrix $\rho$ and two-body pairing tensor $\kappa$.

Inserting the ansatz of Eq.(\ref{eq:gcmApprox}) in the variational principle (\ref{eq:shrodingerVariational}) yields the (time-dependent) Hill-Wheeler equation. In contrast to the static case, there has been no attempt so far to solve the time-dependent Hill-Wheeler equation numerically, as the computational resources needed are beyond current capabilities. Instead, a widespread approach consists in assuming that the norm kernels $\langle\Psi(\qVec)|\Psi(\qVec')\rangle$ can be approximated by a Gaussian form factor \cite{libert_microscopic_1999}. Inserting this Gaussian overlap approximation (GOA) into the Hill-Wheeler equation (using a second order expansion in $\qVec - \qVec' $) leads to a local, time-dependent, Schr\"odinger-like equation in the space $\mathcal{Q}$ of collective coordinates $\qVec$,
\begin{equation} 
\label{eq:evolution0} 
i \hbar \frac{\partial}{\partial t} g(\qVec,t)= \left[ \HCollExpr\right] g(\qVec,t),
\end{equation} where
\begin{itemize}
\item The function $g(\qVec,t)$ is complex. It is related to the weight function $f(\qVec, t)$ appearing in Eq.(\ref{eq:gcmApprox}) and contains all the information about the dynamics of the system. Moreover, the quantity $|g(\qVec,t)|^2$ can be interpreted as the probability density for the system to be in the state $|\Psi(\qVec)\rangle$ at time $t$; see also section \ref{sec:fluxFormalism} below.
\item The real scalar field $V(\qVec)$ and the real symmetric tensor field $B_{kl}(\qVec)$ are fully determined by the knowledge of the effective Hamiltonian $\hat{H}$ and the generator states $|\Psi(\qVec)\rangle $. They reflect the static nuclear properties of the system under study.
\end{itemize}
Throughout this paper, equation~(\ref{eq:evolution0}) will be referred to as the TDGCM+GOA equation. 

\subsection{Collective flux and fission fragment distributions}
\label{sec:fluxFormalism}

The TDGCM+GOA equation implies a continuity equation for the probability density $|g(\qVec,t)|^2$,
\begin{equation}
\label{eq:continuityEquation}
\frac{\partial}{\partial t} |g(\qVec, t)|^2 = -\nabla \cdot \JVec(\qVec, t).
\end{equation}
The real vector field $\JVec(\qVec, t)$ is thus a current of probability. It can be  expressed formally as a function of the collective wave-function,
\begin{equation}
\JVec(\qVec,t)= \frac{\hbar}{2i} B(\qVec) \left[ g^*(\qVec,t) \nabla g(\qVec,t) - g(\qVec,t) \nabla g^*(\qVec,t) \right].
\end{equation}
Specifically, the coordinates of the current of probability read
\begin{equation}
J_{k}(\qVec,t)= 
\frac{\hbar}{2i} \sum_{l=1}^{N} B_{kl}(\qVec) 
\left[ 
g^*(\qVec,t) \frac{\partial g}{\partial q_{l}}(\qVec,t) 
- 
g(\qVec,t) \frac{\partial g^{*}}{\partial q_{l}}(\qVec,t) 
\right].
\label{eq:current}
\end{equation}
As our system evolves in time, its density probability will flow starting from the area of the collective space $\mathcal{Q}$ where the initial wave-function was localized. This evolution is driven by the Hamiltonian $\hat{H}$ through the inertia tensor $B_{kl}(\qVec)$ and the potential energy surface $V(\qVec)$. 

In the case of fission, the potential energy surface is computed up to the points $\qVec$ where the nuclear geometry corresponds to two well-separated fragments. One can thus partition the space $\mathcal{Q}$ into a region where the nucleus is whole -- the internal region, and another where it has split in two fragments -- the external region. The hyper-surface separating the two regions corresponds to the set of scission configurations. The rigorous definition and accurate determination of these scission configurations are themselves challenging problems, which go beyond the scope of this paper; see Refs.~\cite{davies1977,bonneau_microscopic_2007,dubray_structure_2008,younes_microscopic_2009,younes_nuclear_2011,schunck_microscopic_2014} for additional discussions. For practical calculations of fission fragment distributions with \theCode \ we will simply assume the existence of such a scission hyper-surface. 

In general, the local, one-body density matrix $\rho(\rVec)$ in each of the scission points in the collective space $\mathcal{Q}$ is characterized by two high-density regions separated by a thin neck. Assuming the neck is located along the $z$-axis of the intrinsic reference frame, the charge and mass of each fragment can be obtained by simple integration of $\rho(\rVec)$ over the domains $z\in ] -\infty, z_{N}]$ and $z\in [z_{N}, +\infty[$; see, e.g.,  \cite{dubray_structure_2008,younes_microscopic_2009,schunck_microscopic_2014}. According to this procedure, one can associate with each point $\qVec$ of the scission hyper-surface a pair of fragment masses. It follows that the flux of the probability current (\ref{eq:current}) through the scission hyper-surface gives a very good estimate of the relative probability of observing a given pair of fragments at time $t$. We thus define the integrated flux $F(\xi,t)$ through an oriented surface element $\xi$ as
\begin{equation}
F(\xi,t) = \int_{t=0}^{t} dT
       \int_{\qVec\in\xi} 
       \JVec(\qVec,t)\cdot d\SVec.
\label{eq:fluxDef}
\end{equation}
Following \cite{berger_time-dependent_1991, younes_collective_2012}, the fission fragment mass yield for mass $A$ is defined formally as 
\begin{equation}
Y(A) \propto \sum_{\xi\in\mathcal{A}} \lim_{t\rightarrow +\infty} F(\xi, t),
\label{eq:yield}
\end{equation}
where $\mathcal{A}$ is the set of all oriented hyper-surfaces $\xi$ belonging to the scission hyper-surface such that one of the fragments has mass $A$. 
In practice, our calculation of the fragments mass number produces non integer values.
Moreover, one elementary surface $\xi$ may contain several fragmentations.
In this work, we equally distribute the flux component $F(\xi,t)$ between the masses calculated at the vertices of the edge $\xi$:
\begin{equation}
Y(A) = C \sum_{\xi} \frac{1}{N} \sum_{v\in\mathcal{A(\xi)}} \lim_{t\rightarrow +\infty} F(\xi, t),
\label{eq:practicalYield}
\end{equation}
The sum on $\xi$ runs on the whole scission hyper-surface.
The set $\mathcal{A(\xi)}$ contains the vertices of $\xi$ at which one of the fragments has a mass in the interval $[A-1/2; A+1/2]$.	
The normalization constant $C$ is chosen as usual such that
\begin{equation}
\sum_{A=0}^{A_{\text{total}}} Y(A) = 200.
\label{eq:yield_norm}
\end{equation}
In practice, the flux is only integrated from $t=0$ to $t=t_{\text{max}}$. Equations (\ref{eq:current})-(\ref{eq:yield_norm}) show how to extract fission fragment yields from the knowledge of the collective wave function $g(\qVec,t)$ solution to the TDGCM+GOA equations.

\section{Numerical methods}
\label{sec:numericalResolution}

In this section, we detail the numerical methods implemented in the code \theCode \ to solve the TDGCM+GOA equation (\ref{eq:evolution0}) and calculate the flux defined in Eq.~(\ref{eq:fluxDef}).

\subsection{Restriction to a finite domain of space}
\label{subsec:restrictionFiniteDomain}

\theCode \  solves Eq.~(\ref{eq:evolution0}) in a finite domain $\Omega$ of the collective space $\mathcal{Q}$. To ensure the uniqueness of the solution, Dirichlet conditions are imposed at the boundary $\partial \Omega$ of the domain,
\begin{eqnarray}
\label{eq:dirichlet}
\forall \qVec \in \partial \Omega : \quad  g(\qVec , t)= 0.
\end{eqnarray}
Imposing this condition is justified as long as the actual solution $g(\qVec, t)$ is well confined inside the domain $\Omega$ during the whole time evolution of the system. In practice, this may require choosing an excessively large domain $\Omega$. In the case of fission for example, only the internal region discussed in section~\ref{sec:fluxFormalism} and its interface with the external zone present a physical interest. However, we cannot limit $\Omega$ to this area because the probability to observe the fissioning system outside of this configuration subset is not negligible.

To circumvent this issue, \theCode \  defines an absorption band along the boundary $\partial \Omega$. This band artificially simulates the leakage of the wave packet $g(\qVec, t)$ outside of the calculation domain. Formally, absorption is taken into account by introducing a new imaginary term in the evolution equation,
\begin{multline}
\label{eq:evolution1} 
\forall \ \qVec \in \Omega, \  t \in [0,t_{\text{max}}]:\\
i \hbar \frac{\partial}{\partial t} g(\qVec,t)= \left[ \HCollExpr -i\hbar A(\qVec) \right]g(\qVec,t).
\end{multline} 
The real scalar field $A(\qVec)$ is non zero only in the absorption band. In this region, $A(\qVec)$ is taken as a simple polynomial increasing smoothly from 0 on the inner border of the band and reaching its maximum at the boundary of the domain,
\begin{equation}
A(\qVec)= 4r\left(1-\frac{x(\qVec)}{w}\right)^3.
\end{equation}
The quantity $x(\qVec)$ is the minimal Euclidean distance between the point $\qVec$ and the boundary $\partial \Omega$. The parameters $r$ and $w$ correspond to the average absorption rate and width of the absorption band respectively. These two parameters can be tuned by the user as a function of the problem characteristics to ensure optimal absorption.

\subsection{Space discretization }
\label{tdgcm_solving_spaceDiscretization}

As mentioned earlier, we use the Galerkin finite element method \cite{chen_finite_2005,brenner_mathematical_2008} to discretize the collective space $\mathcal{Q}$. The main reasons for choosing this approach are its capability to manage non regular meshes and the possibility to apply h-refinement and p-refinement techniques to improve computational efficiency. In this section, we show how to formally derive a linear system of equation from the discretization of Eq.~(\ref{eq:evolution1}).

As customary in quantum mechanics, we note $\langle . |  . \rangle$ the scalar product in the space $ \mathcal{L}^2(\Omega, \mathbb{C})$ of complex-valued, square-integrable functions,
\begin{equation}
\langle \phi |\psi\rangle = \int_\Omega d\qVec\; \phi^*(\qVec) \,\psi(\qVec).
\end{equation}
With this definition, Eq.~(\ref{eq:evolution1}) can be recast into
\begin{equation}
\forall \phi \in \mathcal{L}^2(\Omega, \mathbb{C}),\ 
 \forall t \in [0,t_{\text{max}}]: \quad 
 \langle \phi | r(t) \rangle = 0,
\end{equation}
with the residual $r(\qVec, t)$ defined as
\begin{multline} 
r(\qVec, t)=  \left[ \HCollExpr   \right. \\
\left. -i\hbar A(\qVec) - i\hbar\frac{\partial}{\partial t} \right] g(\qVec,t)
\end{multline}

Following the standard approach of the finite elements method, the domain $\Omega$ is first partitioned into a mesh. In our case, each cell of the mesh is a $N$-dimensional simplex (triangle if $N=2$, tetrahedron if $N=3$, etc.). We note $\mathcal{S}$ the set of all simplices in the domain. Inside every simplex of the mesh, we assume a polynomial form for the numerical solution of Eq.~(\ref{eq:evolution1}). At any time $t$ and in any simplex $s\in\mathcal{S}$, we thus define the local interpolating polynomial $P_{s,t}$
\begin{equation} 
\forall s\in\mathcal{S}, \forall \qVec \in s: \, g(\qVec,t)=P_{s,t}(\qVec).
\end{equation} 
For each simplex $s\in\mathcal{S}$, we select the degree $d_s$ of the interpolating polynomial. The space $\mathcal{P}_{s}$ of all interpolating polynomials in the simplex $s\in\mathcal{S}$ is a vector space. Its dimension $D_{s}$ is given by the binomial coefficient,
\begin{equation}
D_{s} = \left( \begin{array}{c} N+d_{s} \\ d_{s} \end{array}\right).
\end{equation}
In order to discretize Eq.~(\ref{eq:evolution1}), we now build a convenient basis of the space $\mathcal{P}_{s}$. First, we define for each simplex $s\in\mathcal{S}$ a finite set of specific points $\qVec_{j}\in s$ called nodes. Next, we introduce a set of real polynomials $\phi_{s,i}$ associated with the simplex $s$. For all nodes $i$ of the simplex $s$, the polynomial $\phi_{s,i}$ is defined by the requirement
\begin{equation}
\label{eq:localBasis}
\phi_{s,i}(\qVec_j)= \left \{ 
\begin{array}{c} 
1, \quad \text{if } i=j \\ 0, \quad \text{if } i \neq j \\
\end{array} 
\right . 
\end{equation}
In other words, the $\phi_{s,i}$ are the usual Lagrange polynomials. The total number of nodes in each simplex $s\in\mathcal{S}$ is equal to $D_{s}$, so that the set  $\{\phi_{s,i}\}_{i=1,D_{s}}$ forms a basis of $\mathcal{P}_{s}$. The total number of nodes in the entire domain $\Omega$ is noted $m$.

With the help of these local bases, we can define for each of the $m$ nodes $i$ of the domain $\Omega$ a function $\phi_i$ such that
\begin{equation} 
\forall \qVec \in \Omega: \quad \phi_{i}(\qVec)= \left \{ 
\begin{array}{ll} 
\phi_{s,i}(\qVec), & \text{if } \qVec \in s \text{ and } i \text{ is a node of } s \\ 
0, & \text{otherwise} \\ 
\end{array} \right . 
\end{equation} 
The functions $\{\phi_{i}\}_{i=1,m}$ form a basis of our solution space. The solution of  Eq.~(\ref{eq:evolution1}) can thus be expanded as
\begin{equation}
\label{eq:gExpand}
g(\qVec,t)= \sum_{ i=1}^m g(\qVec_i,t)\phi_i(\qVec).
\end{equation}

Applying the Galerkin finite element method, we search for a numerical solution $g(\qVec,t)$ of the form (\ref{eq:gExpand}) that verifies
\begin{equation} 
\forall t \in [0,t_{\text{max}}], \forall i \in [1,m]: \quad
\langle \phi_i(\qVec) | g(\qVec,t) \rangle = 0. 
\end{equation} 
This process yields a discretized system of $m$ equations with the $m$ coefficients $g(\qVec_j,t)$ as the unknown. It can be written in the condensed form
\begin{equation} 
\label{eq:spaceDiscretizedEvolution}
i\hbar M \, \frac{\partial G(t)}{\partial t}= [H -i\hbar A]G(t),
\end{equation}
where $G(t)$ denotes the $m$-dimensional vector of coefficients $g(\qVec_j,t)$ at every node $j$ of the domain $\Omega$. The $m\times m$ matrices $ M, H$ and $A$ are defined by 
\begin{equation} 
\begin{array}{l}
M_{ab}= \langle\phi_a(\qVec) | \phi_b(\qVec) \rangle, \medskip\\ 
A_{ab}= \langle\phi_a(\qVec) | A(\qVec)\phi_b(\qVec)\rangle, \medskip\\ 
H_{ab}= \langle\phi_a(\qVec) | \left[ \displaystyle\HCollExpr\right]\phi_b(\qVec)\rangle.
\end{array}
\end{equation}
All matrix elements can be computed by applying the basis expansion (\ref{eq:gExpand}) to the fields $V(\qVec)$, $B_{kl}(\qVec)$ and $A(\qVec)$, that is, 
\begin{equation}
\label{eq:fieldExpand}
F(\qVec)= \displaystyle\sum_{c=1}^m F(\qVec_c)\,\phi_c(\qVec) \quad \text{with } F=V,\, B_{kl},\text{ or }A.
\end{equation} 
The double derivative term in $H_{ij}$ can be integrated by parts using the Dirichlet conditions imposed on the boundary $\partial \Omega$. The final expression for the matrix elements is
\begin{equation}
\label{eq:matrixElements}
\begin{array}{rl} 
M_{ab}= & \displaystyle \int_\Omega d\qVec\; \phi_a(\qVec)\,\phi_b(\qVec) \medskip\\ 
A_{ab}= & \displaystyle \sum_{c=1}^m A(\qVec_c) \int_\Omega d\qVec\;  \phi_a(\qVec)\,\phi_b(\qVec)\,\phi_c(\qVec) \medskip\\
H_{ab}= & \displaystyle \sum_{c=1}^m V(\qVec_c) \int_\Omega d\qVec\;  \phi_a(\qVec)\,\phi_b(\qVec)\,\phi_c(\qVec) \\
        & \displaystyle -\sum_{c=1}^m \sum_{kl} B_{kl}(\qVec_c) \int_\Omega d\qVec\;  \frac{\partial \phi_a}{\partial q_k}(\qVec) \frac{\partial \phi_b}{\partial q_l} (\qVec)\, \phi_c(\qVec).
\end{array} 
\end{equation}
Since the basis functions $\phi_{a}(\qVec)$ are simple polynomials of $\qVec$, integrations can be performed analytically, and no quadrature or numerical integration scheme is needed. Note that the three matrices obtained here are real, symmetric and sparse. The sparsity comes from the fact that the overlap between any two basis functions is zero unless at least one element was defined with both the corresponding nodes.

The values of each field at each nodes are inputs of the calculation. To compute the matrix elements, \theCode \ relies on a formal representation of the polynomials. Basis elements $\phi_{s,i}$ are first derived from Eq.~(\ref{eq:localBasis}). This step requires inverting a small dense linear system for each simplex. Then derivatives, multiplications and integrations of the polynomials involved in (Eq.~\ref{eq:matrixElements}) can all be performed formally, so that these operations do not generate other errors than those related to the accuracy of the polynomial coefficients .

\subsection{Time discretization }
\label{tdgcm_solving_timeDiscretization}

The Crank-Nicolson scheme is used to discretize Eq.~(\ref{eq:spaceDiscretizedEvolution}) in time \cite{crank_practical_1996}. We recall that the Crank-Nicolson scheme gives the following prescriptions for the function and its time-derivative,
\begin{equation}
\frac{\partial G}{\partial t} \simeq \frac{G(t+\Delta t)- G(t)}{\Delta t}, 
\qquad 
G(t) \simeq \frac{G(t+\Delta t) + G(t)}{2}.
\end{equation}
Starting from Eq.~(\ref{eq:spaceDiscretizedEvolution}), this numerical scheme yields the fully discretized equation
\begin{equation} 
\label{eq:fullyDiscretized}
R \times G(t+\Delta t) = b(t),
\end{equation}
with
\begin{equation}
\begin{array}{l}
R = \displaystyle M + \frac{\Delta t}{2}A + i\frac{\Delta t}{2\hbar}H,
\medskip\\ 
b(t)= \displaystyle \left[ M - \frac{\Delta t}{2}A - i\frac{\Delta t}{2\hbar}H \right]  G(t).
\end{array}
\end{equation}

Using this time discretization scheme, we can show that both the norm of the collective wave-function and the average energy computed from the numerical solution are both constants in time if the absorption term is set to zero
\begin{equation}
\label{eq:normConst}
\begin{array}{l}
||g(\qVec,t)||_2= \displaystyle \left[ \int_\Omega g^*(\qVec,t)  g(\qVec,t) \right ]^{1/2}= \text{cst}, \\
\\
\displaystyle \int_\Omega g^*(\qVec,t) \left[\HCollExpr\right] g(\qVec,t) =\text{cst}.
\end{array}
\end{equation}
These properties can be used to test the validity of the numerical implementation.

\subsection{Inversion of a linear system}
\label{sec:linearSystemInversion}

The collective wave-function at time $t$ is obtained by solving the fully-discretized  Eq.~(\ref{eq:fullyDiscretized}) for the vector $G(t)$. This requires inverting at each time step a complex, sparse $m\times m$ matrices. In \theCode \ these inversions are computed with the iterative QMR algorithm without look-\/ahead as described in \cite{freund_conjugate_1992}. The numerical solution at iteration $n$ is used as the initial guess for iteration $n+1$. The convergence criterion to stop the iterations for each inversion is defined by
\begin{equation} 
\label{eq:convergenceCriteria}
||R \times G(t+\Delta t) - b ||_2 < \epsilon || b ||_2,
\end{equation}
with the tolerance $\epsilon$ specified by the user. In order to accelerate the inversion, a Jacobi preconditioner is applied to the system before the first time iteration.

Since no look-\/ahead statement is implemented, the QMR algorithm may occasionally fail to converge at the level of precision required. In such case, the system is rewritten in the $(2m \times 2m)$ real form
\begin{equation}
\left(
\begin{array}{c c} 
M + \frac{\Delta t}{2}A & -\frac{\Delta t}{2\hbar}H \\ 
\frac{\Delta t}{2\hbar}H & M + \frac{\Delta t}{2}A \end{array} \right) 
\left( \begin{array}{c} 
\mathfrak{Re}(G(t+\Delta t)) \\ 
\mathfrak{Im}(G(t+\Delta t)) 
\end{array} \right) 
= 
\left( \begin{array}{c}
\mathfrak{Re}(b(t)) \\ 
\mathfrak{Im}(b(t))
\end{array} \right),
\end{equation} 
where $\mathfrak{Re}$ and $\mathfrak{Im}$ refer to the real and imaginary parts, respectively. This real system is then solved with the Bi-conjugate Gradient Stabilized Method as described in \cite{saad_iterative_2003}. 

In practice, the small numerical errors caused by these matrix inversions accumulate over time, and can lead to violations of the properties (\ref{eq:normConst}). However, this numerical error is very small, especially if the time span of the time iterations is reasonable and the tolerance $\epsilon$ is small enough.

\subsection{Calculation of the flux}
\label{sec:fluxCalculation}

In \theCode, hyper-surfaces are defined as the union of oriented faces of an arbitrary list of simplices in the mesh. Note that the hyper-surfaces thus defined are not necessarily connected, as the simplices need not be adjacent. Given such a hyper-surface provided by the user, the code  can compute the flux $F(\xi)$ as defined in Eq.~(\ref{eq:fluxDef}) through each of the faces $\xi$. 

The instantaneous elementary flux $f$ going through an oriented simplex face $ \xi $ at time $t$ is calculated as 
\begin{equation}
\label{eq:fluxInst}
\text{f}(\xi, t)= \sum_{k=1}^N n^{(\xi)}_{k}\cdot \int_{\qVec \, \in\, \xi} J_k(\qVec,t)  \, dS
\end{equation} 
where $\boldsymbol{n}^{(\xi)}$ is the unit vector normal to the simplex face $ \xi $, $\JVec(\qVec,t)$ is the probability current (\ref{eq:current}) and $N$ is the dimension of the collective space
\footnote{This definition is only valid if the collective space has a dimension strictly superior to one. The flux calculation is not enabled in \theCode \ in the case of 1-dimensional spaces. } .
In order to compute the instantaneous flux, we expand the collective wave function $g(\qVec,t)$ and the inertia tensor field $B_{kl}(\qVec)$ on the FE basis using Eqs.~(\ref{eq:gExpand})-(\ref{eq:fieldExpand}). The integral in the flux becomes
\begin{multline} 
\int_{\qVec \, \in\, \xi} J_k(\qVec,t) \, dS 
= 
\hbar \sum_{l=1}^N \sum_{u,v,w} B_{kl}(\qVec_{u}) \\
\times \left[ 
\mathfrak{Re}(G_{v}(t))\mathfrak{Im}(G_{w}(t)) 
- 
\mathfrak{Im}(G_{v}(t))\mathfrak{Re}(G_{w}(t)) \right] 
 I_{\xi,u,v,w,l},
\end{multline}
with
\begin{equation}
\label{eq:fluxIntegral}
 I_{\xi,u,v,w,l}=\int_{\qVec \, \in\, \xi} \phi_u(\qVec)\,\phi_v(\qVec)\frac{\partial \phi_w(\qVec)}{\partial q_l} dS .
\end{equation}
The integral $I_{\xi,u,v,w,l}$ is non-zero only if $u$, $v$ and $w$ are nodes of a same simplex containing the edge $\xi$.
For these non-zero terms, the integration is performed formally. This is achieved by first computing the polynomial expression of the integrand. Then, two successive changes of variables are applied to reduce the domain of integration to a simplex of dimension $N-1$. Finally, we use a trapezoid rule to integrate over time the instantaneous flux $f(\xi,t)$ to obtain the expression Eq.~(\ref{eq:fluxDef}).

\begin{figure}[!ht]
  \begin{center}
  \includegraphics[width=0.20\textwidth]{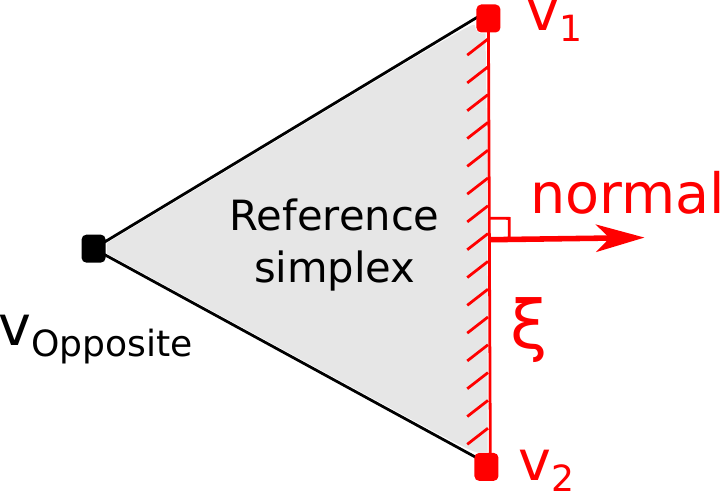}
  \caption{Definition of the reference simplex for a 2-dimensional mesh}
  \label{fig:vOpposite}
  \end{center}
\end{figure}

The numerical methods presented here do not enforce the continuity of the derivative of the solution at the interface $\xi$. Therefore, the integral $I_{\xi,u,v,w,l}$ may not always be defined: for two simplices $s$ and $s'$ sharing the interface $\xi$, the value of the partial derivative $\partial g/\partial q_{l}$ at any point $\qVec\in\xi$ may be different if computed from the expansion in the basis $\mathcal{P}_{s}$ or in the basis $\mathcal{P}_{s'}$. In practice, only two simplices share the interface $\xi$. In \theCode\ the value of the integral $I_{\xi,u,v,w,l}$ is computed from one of these two simplices, which we call the reference simplex. By convention, we define the reference simplex as opposite to the direction of the normal unit vector, as illustrated in figure~\ref{fig:vOpposite}.

\section{Benchmarks}
\label{sec:benchmarks}

In this section, we present a series of benchmark calculations that highlight specific features of the code. In each case, the analytical solution verifies one or several properties that we use to test our numerical implementation. To this purpose, we define for each case the error between the numerical solution and the analytical result, and compute this error as a function of the numerical parameters of our calculation, namely
\begin{itemize}
 \item the numerical tolerance $\epsilon$ for matrix inversions; see Eq.~(\ref{eq:convergenceCriteria});
 \item the mesh size $h$, which provides an estimate of the ``spatial'' resolution of the domain $\Omega$;
 \item the degree $d_{s}$ of basis polynomials;
 \item the time step $\delta t$ used in time integration.
\end{itemize}

\subsection{Conservation of the Norm}
\label{subsec:conservationNorm}

The conservation of the norm expressed by Eq.(\ref{eq:normConst}) is the simplest test of our implementation. In this benchmark, the calculation domain $\Omega$ is a 3D cube of size 10 arbitrary units (a.u.). The mesh is built by creating a regular grid of equidistant vertices with a mesh size $h=1$ a.u.. The position of each vertex is then randomly perturbed. A new coordinate $q'_k$ is sampled uniformly in the interval $[q_k-f_h h; q_k+f_h h]$, where $f_h$ is a fluctuation factor set to 15\%, and $q_k$ is the associated old coordinate. Once the vertices are defined, the simplices are determined by Delaunay triangulation. In each simplex, a polynomial basis of degree $d_{s} =1 $ is used and the nodes are positioned exactly at the vertices.

Based on this mesh, a free wave packet is propagated during a time $t_{\text{max}}=15$ a.u.. The inertia tensor is diagonal, inversely proportional to a mass $m=1.3$ a.u. and independent of space. The initial wave packet is a 3D isotropic Gaussian centered in the middle of the simulation domain and characterized by the width $\sigma= \sqrt{\hbar/1.04}$ a.u.. Finally, we use the time step $\delta t=5.10^{-4}$ a.u., and the numerical tolerance for matrix inversions is set to its default value of $10^{-15}$.

\begin{figure}[!ht]
  \begin{center}
  \includegraphics[width=0.45\textwidth]{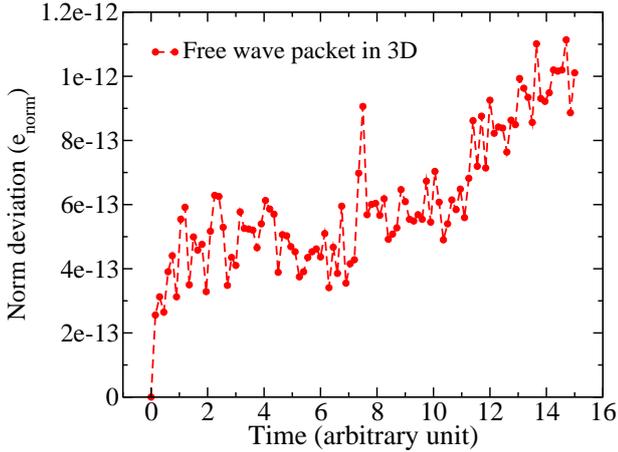}
  \caption{Error of the norm $e_{\text{norm}}$ as a function of time for the free isotropic Gaussian wave-packet.}
  \label{fig:normDeviation}
  \end{center}
\end{figure}

The error $e_{\text{norm}}$ of the norm is defined as
\begin{equation}
e_{\text{norm}}= \frac{| \, ||g(\qVec,t)||_2-||g(\qVec,0)||_2 \,|}{||g(\qVec,0)||_2}
\end{equation}
This quantity is computed at different times of the simulation and is plotted in figure~\ref{fig:normDeviation}. We note that the error is maintained below $10^{-12}$ during the whole simulation with this choice for the numerical tolerance of matrix inversions. This is consistent with the expected property of Eq.~(\ref{eq:normConst}). As discussed in section~\ref{sec:linearSystemInversion} the error $e_{\text{norm}}$ comes from the accumulation of errors from matrix inversions at each time step, which explains its increase with time.

In all subsequent calculations discussed in this paper, we use the same numerical tolerance of $10^{-15}$ for matrix inversion and the error $e_{\text{norm}}$ on the norm will always be below $10^{-9}$. With this level of numerical precision, errors coming from matrix inversion will always be several orders of magnitude below any other source of numerical error examined in this work.

\subsection{Harmonic oscillator potential}
\label{subsec:harmonicOscillator}

We now turn to the dynamics of a quantum system in an isotropic harmonic oscillator (HO) potential in $N=1$ and $N=2$. The advantage of the HO potential is that it provides analytical solutions that can be used to test the implementation.

\begin{table}[!ht]
\begin{center}
 \begin{tabular}{cccc} 
 \hline
 m   & $\omega$ &$\Omega$      &$f_h$ \\
 \hline
 1.3 & 0.8      & [-20;20]$\displaystyle^N$ & 0.15 \\
  \hline
 \end{tabular}
 \caption{Characteristics of the 1D and 2D harmonic oscillators used in this study.}
 \label{tab:hoCharacteristics}
 \end{center}
\end{table}

In the following calculations, the inertia tensor is always diagonal, inversely proportional to a mass $m$ and independent of space, $B_{kl}(\qVec)=\delta_{kl}/m$. The HO potential being isotropic, it is characterized by a single frequency $\omega$,
\begin{equation}
V(\qVec)= \frac{1}{2} m\omega \qVec^{2}.
\end{equation}
The numerical values adopted for the HO potential are listed in table~\ref{tab:hoCharacteristics}. All calculations are performed in a domain $\Omega=[-20;20]^N$. The same procedure as described in section~\ref{subsec:conservationNorm} is used to build an irregular mesh. Dirichlet boundary conditions are enforced at the boundaries $\partial\Omega$ of the domain. On the other hand, we find that the analytic expressions for the first two eigenstates of the HO verify
\begin{equation}
\forall \qVec \in \partial \Omega: \quad 
 \frac{|g(\qVec)|}{\text{max}\{ \, |g(\qVec)|\, \} }_{\qVec \in \Omega } < 2.10^{-13}.
\end{equation}
Therefore, the numerical error coming from the finite size of the domain of resolution $\Omega$ is completely negligible compared to the other sources of error under study.

\subsubsection{Ground state of the 1D HO}
\label{sec:fundamentalState}

For any eigenstate of the potential, the modulus $|g(\qVec, t)|$ of the wave function is independent of time. We first test this property for the ground state of the 1D HO, $\qVec \equiv q$. The initial wave function is taken as
\begin{equation}
 g(Q,t=0)= \operatorname{exp}\left(-\frac{Q^2}{2}\right),
\end{equation}
where the reduced coordinate $Q$ is defined as
\begin{equation}
 Q = \sqrt{\frac{m\omega}{\hbar}} q.
\end{equation}
Calculation is performed up to $t_{\text{max}}=32$ a.u., which is slightly larger than two periods of the complex function $g(\qVec,t)$. The deviation of the modulus of the numerical solution at time $t$ from its initial value is measured by the error $e_{\text{mod.}}$,
\begin{equation}
 e_{\text{mod}}= \frac{\| \, |G(t_{\text{max}})|- |G(t=0)| \, \|_\infty}{\| \,  |G(t=0)| \, \|_\infty},
\end{equation}
where the infinity norm $||\dots||_{\infty}$ for a vector $G$ refers to the maximum absolute value of its elements. This error has been computed from a set of calculations with different mesh sizes $h$ and time steps $\delta t$. Results are presented in  table~\ref{tab:ho1dStatic}.

\begin{table}[!ht]
\begin{center}
$
 \begin{array}{c|ccc} 
 \hline

      \delta t\,  | \,h     & 1.0  & 0.1 & 0.01 \\
 \hline
 10^0        &  8.089.10^{-4}    & 1.039.10^{-5}   & 7.460.10^{-7}     \\
 10^{-1}     &  3.780.10^{-3}    & 7.616.10^{-6}   & 1.708.10^{-6}     \\
 10^{-2}     &  3.688.10^{-3}    & 2.603.10^{-5}   & 2.555.10^{-6}    \\
 10^{-3}     &  3.662.10^{-3}    & 2.626.10^{-5}   & 2.461.10^{-6} \\
 10^{-4}     &  3.661.10^{-3}    & 2.619.10^{-5}   & 2.603.10^{-6} \\
 \hline
 \end{array}
 $
 \caption{Error $e_{\text{mod}}$ on the modulus $|g(q,t)|$ for the ground-state of a 1D HO.}
 \label{tab:ho1dStatic}
 \end{center}
\end{table}
 
We verify that the error $e_{\text{mod}}$ decreases when refining the calculation in both time and space. For a given mesh size $h$, the error slightly fluctuates with $\delta t$ before it finally reaches a converged value. The major part of the error is clearly driven by space discretization.

\subsubsection{Sum of two eigenstates for the 1D HO}

In this benchmark, the initial state of the system is a sum of the first two eigenstates of the 1D HO, and can thus be written
\begin{equation}
 g(Q,t=0)= \operatorname{exp}\left(-\frac{Q^2}{2}\right) (1+Q).
\end{equation}
The second eigenstate is characterized by a frequency three times larger than in the ground-state. Therefore, the motion is periodic with the same period as in section~\ref{sec:fundamentalState}. In this case, the modulus of the wave function is not constant anymore, since the system is not in an eigenstate, but it oscillates between two positions. The full time-dependent solution is given analytically by
\begin{equation}
 g(Q,t)= \operatorname{exp}\left(-\frac{Q^2}{2}\right) e^{-i\omega t/2}
 \left(1+Q e^{-i \omega t} \right).
\end{equation}

\begin{figure}[h]
  \begin{center}
  \includegraphics[width=0.45\textwidth]{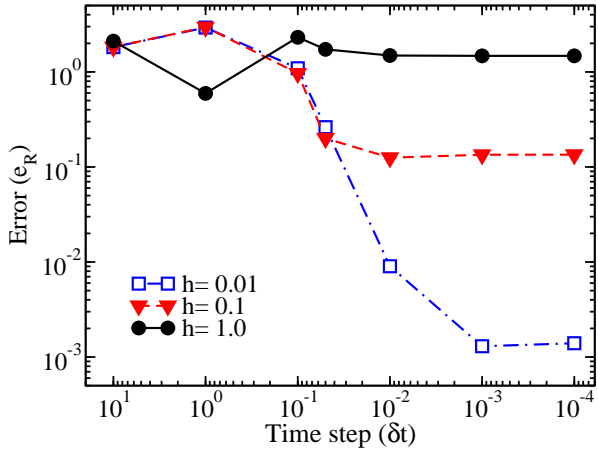}
  \caption{Error $e_{R}$ on the real part of the solution for a periodic motion in a 1D HO.}
  \label{fig:ho1dPerodic}
  \end{center}
\end{figure}

We compare the real part of the numerical solution $\mathfrak{Re}(g)(\qVec,t)$ to its analytic expression using the error $e_{R}$ defined as
\begin{equation}
 e_{R}= \frac{|| \mathfrak{Re}(g_{\text{num.}}(t))- \mathfrak{Re}(g_{\text{the.}}(t)) ||_\infty}{|| \mathfrak{Re}(g_{\text{the.}}(t)) ||_\infty}.
\end{equation}
This error is computed for several time and space refinements and is plotted as a function of time step in figure~\ref{fig:ho1dPerodic}. As in the previous benchmark, we observe that the numerical solution converges to its analytic value as time and space are refined.

\subsubsection{Sum of two eigenstates in 2D}
\label{sec:sumTwoEigenStates}

In this section, the previous study is generalized to the case of a 2D HO, $\qVec \equiv (q, q')$. This benchmark of a 2D case allows us to compare the numerical calculation of the flux with its analytic expression. The initial wave function reads
\begin{equation}
 g(Q,Q',t=0)= \operatorname{exp}\left(-\frac{Q^2+Q'^2}{2}\right) (1+Q),
\end{equation}
where $Q'$ is the reduced coordinate associated with $q'$. Starting from this state, the full time-dependent solution reads
\begin{equation}
  g(Q,Q',t)= \operatorname{exp}\left(-\frac{Q^2+Q'^2}{2}\right) e^{-i\omega t}
  \left(1+Q e^{-i \omega t}\right).
\end{equation}
The system oscillates from one side of the line $q=0$ to the other side with a period 2$\pi/\omega\simeq 31.4$ a.u. The derivatives of the wave function read 
\begin{equation}
 \begin{array}{l}
  \displaystyle\frac{\partial g}{\partial q} 
  = 
  \sqrt{\frac{m\omega}{\hbar}} 
  e^{-\frac{Q^2+Q'^2}{2}} e^{-i \omega t} \left(-e^{-i \omega t} Q^2 - Q + e^{-i \omega t}\right), \medskip\\
  \displaystyle\frac{\partial g}{\partial q'}
  = 
  - \sqrt{\frac{m\omega}{\hbar}} e^{-\frac{Q^2+Q'^2}{2}} e^{-i \omega t} \cdot Q' \left( 1 + Q e^{-i \omega t} \right).
 \end{array}
\end{equation}

\begin{figure}[!ht]
 \begin{center}
 \includegraphics[width=0.10\textwidth]{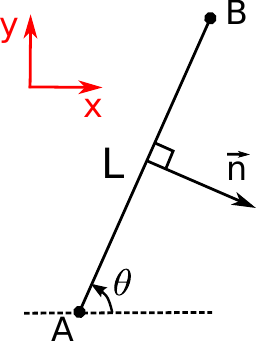}
 \caption{Definition of the segment [A,B]}
 \label{fig:segment}
 \end{center}
\end{figure}

The instantaneous flux of Eq.~(\ref{eq:fluxInst}) through an oriented segment [AB], as depicted in figure \ref{fig:segment}, is given by
\begin{multline}
 f([AB],t)= - \displaystyle \frac{\sqrt{\pi} \, \hbar \, \operatorname{sin}(\theta) \operatorname{sin}(\omega t)}{2 m} 
 e^{-\left( Q_{A}\operatorname{sin}\theta - Q'_{A}\operatorname{cos}(\theta) \right)^2} \\
 \times \Big[ 
 \operatorname{erf}\left(z + Q_A\operatorname{cos}\theta + Q'_A \operatorname{sin}\theta  \right)
 \Big]_{z=0}^{\sqrt{ \frac{m\omega}{\hbar}}L} ,
\end{multline}
with the error function defined as as usual by
\begin{equation}
 \operatorname{erf}(z)= \frac{2}{\sqrt{\pi}}\int_0^z e^{x^2} dx.
\end{equation}

\begin{figure}[!ht]
 \begin{center}
 \includegraphics[width=0.45\textwidth]{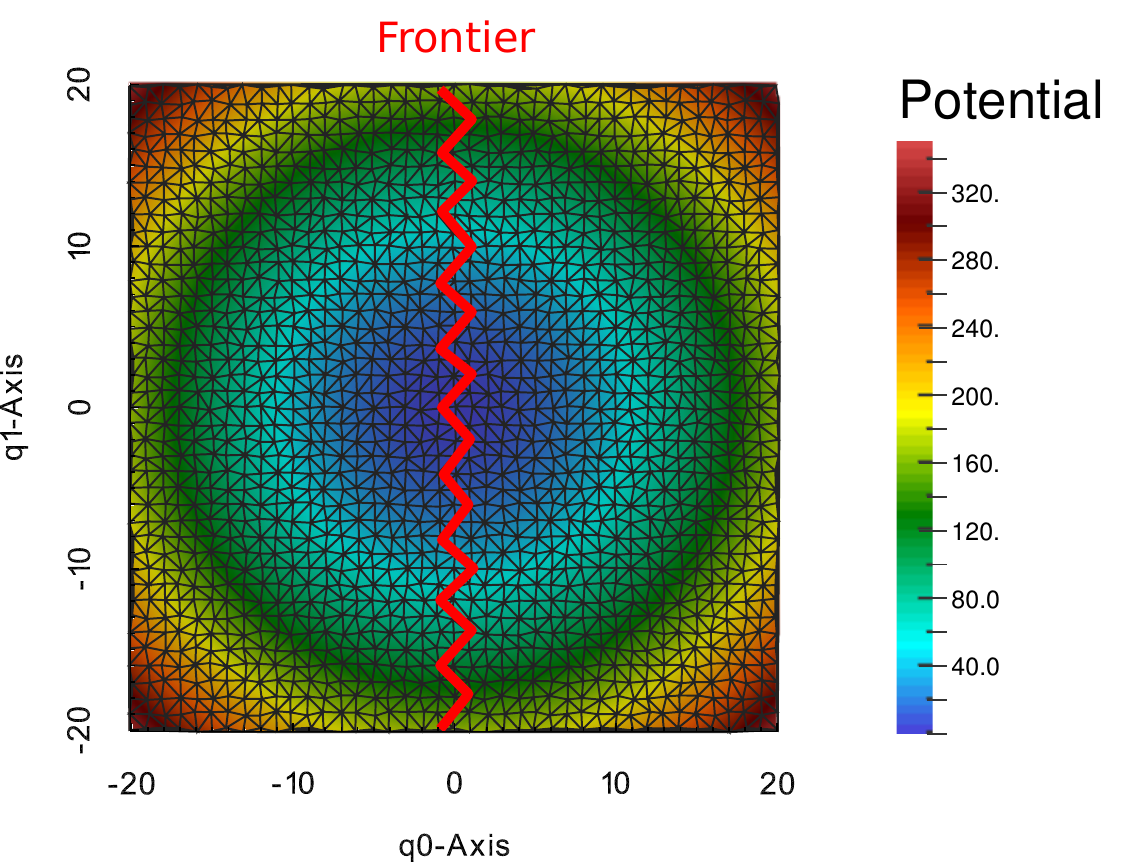}
 \caption{Domain of resolution $\Omega$ for the 2D HO.}
 \label{fig:ho2dMesh}
 \end{center}
\end{figure}

To test our implementation, we first define an arbitrary frontier following an oscillating path around the line $q=0$; see figure \ref{fig:ho2dMesh}. The dynamics of the system is computed up to $t_{\text{max}}=16$ a.u. At each vertex of the frontier, we calculate the quantities $\mathfrak{Re}(g)$ and $\partial\mathfrak{Im}(g)/\partial q$ at the end of the time evolution. In addition, our code provides the numerical value of the instantaneous flux through each element of the frontier. These three vectors of results obtained at the frontier are compared to their respective analytic expressions based on the error
\begin{equation}
 e_{\text{front}}(\vVec) = \frac{|| \vVec_{\text{num.}}- \vVec_{\text{the.}} ||_\infty}{|| \vVec_{\text{the.}} \|_\infty}.
\end{equation}
These errors have been estimated for a time step $\delta t=5.10^{-4}$ and for the following set of different space discretizations,
\begin{itemize}
 \item $h=2, 1, 0.5, 0.2, 0.1$, with a polynomial basis of degree $d_{s}=1$. Decreasing the mesh size $h$ is known as $h$-refinement in the FE approach;
 \item $h=2, 1, 0.5$, with a polynomial basis of degree $d_{s}=2$. For these calculations, each simplex has 6 nodes, of which three are positioned at the vertices and the three others in the middle of each edge.
\end{itemize}
Note that for a same mesh size $h$, going from linear to quadratic polynomials doubles the number of nodes in the mesh. This is an example of p-refinement in the FE approach.

\begin{figure}[!ht]
  \begin{center}
  \includegraphics[width=0.4\textwidth]{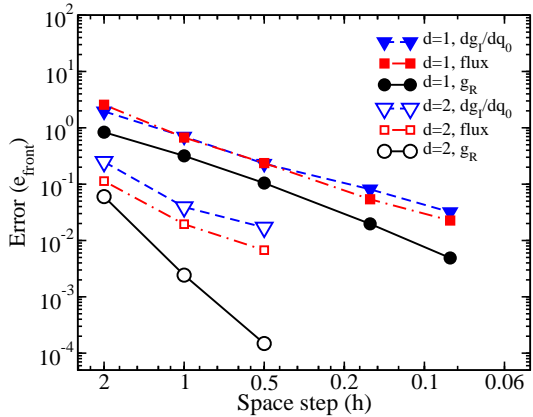}
  \caption{Numerical errors for the real part of the solution, $\mathfrak{Re}(g)$, the spatial derivative of the imaginary part of the solution, $\partial\mathfrak{Im}(g)/\partial q$, and the instantaneous flux as a function of the mesh size $h$ for an isotropic 2D HO. }
  \label{fig:ho2dPeriodic}
  \end{center}
\end{figure}

The results obtained are summarized in figure~\ref{fig:ho2dPeriodic}. Although the results are only presented for one value of the time step $\delta t$, we obtained similar results with another series of calculations using $\delta t=1.10^{-4}$. For the three quantities tested, the convergence to their analytic expression is numerically confirmed. We note that the convergence is much faster when using polynomials of degree two in the basis. Moreover, the error associated with the derivative of the solution is much larger than the one associated with the solution itself. In particular, it seems that the convergence rate of the flux is limited by the error on the derivative.

\subsection{Stokes theorem}

The goal of this benchmark is to test the calculation of the flux in a case where no analytic expression of the solution is available. This will be achieved by the use of the Stokes theorem of differential geometry. As is well known, if we define an enclosed volume $V \subset \Omega$, the continuity equation (\ref{eq:continuityEquation}) yields the following conservation relation
\begin{equation}
\label{eq:stockesTheorem}
e_{Stokes}=
\int_{t=0}^{t_{\text{max}}}dt \int_{\qVec\in \partial V} \JVec(\qVec,t) d\SVec
-
\left[\int_{\qVec \in V} |g(\qVec,t)|^2 d\qVec \right]^{t_{\text{max}}}_{t=0} = 0.
\end{equation}
We tested this property in the case of a 2-dimensional free wave packet. The simulation domain, the mesh, the frontier, and the inertia are the same as in section~\ref{sec:sumTwoEigenStates}. The initial state is a Gaussian function centered at $\qVec=(-5;0)$ and having a width $\sigma=\sqrt{\hbar}/1.04$ a.u. This function is then multiplied by a plane wave of impulsion $\boldsymbol{k}=(1;0)$ and normalized to one.

We define the volume $V$ as one half of $\Omega$ delimited by the frontier $\mathcal{F}$ and containing the major part of the initial wave packet. In this configuration, the Dirichlet condition imposed on the boundary of the simulation domain imposes
\begin{equation}
    \int_{\qVec\in \partial V}  \JVec(\qVec,t)d\SVec = \sum_{\xi \in \mathcal{F}} F(\xi).
\end{equation}
The error $e_{\text{Stokes}}$ on Eq.~(\ref{eq:stockesTheorem}) can therefore be computed at any time from the flux and the numerical solution produced by \theCode.

\begin{figure}[!ht]
  \begin{center}
  \includegraphics[width=0.4\textwidth]{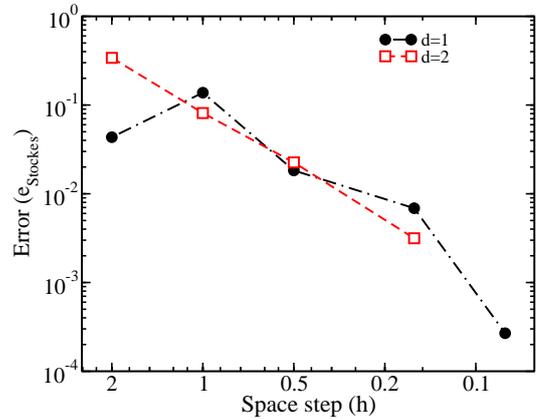}
  \caption{Absolute deviation from the Stokes theorem $|e_{\text{Stokes}}|$}
  \label{fig:stockesThm}
  \end{center}
\end{figure}

In this benchmark, we run the simulation up to $t_{\text{max}}=3$ a.u.. At this time, most of the wave packet has crossed the frontier. The calculation is performed with a time step $\delta t=10^{-4}$ and repeated for several space discretizations. The results are shown in figure \ref{fig:stockesThm}. The property (\ref{eq:stockesTheorem}) is verified up to 0.1\% of the norm.

\section{Application to the fission of \Pu240}
\label{sec:fissionPu240}

To demonstrate the capability of \theCode, we show in this section the results 
obtained by solving the TDGCM+GOA equation for the neutron-induced fission of a 
\Pu239 target. In particular, we emphasize the convergence of such a calculation 
in a realistic setting as well as several features used to increase computational 
efficiency.

\subsection{Description of the calculation}
\label{sec:Pu240Physics}

We solve the TDGCM+GOA equation in the 2D collective space spanned by the 
quadrupole ($q_{20}$) and octupole ($q_{30}$) moments of the fissioning system. 
In the following, $q_{20}$ and $q_{30}$ are expressed in fm$^2$ and fm$^{3}$, 
respectively. The potential energy surface $V(\qVec)$ of the compound nucleus 
\Pu240 in this collective space is obtained by solving the HFB equations with a 
Skyrme energy density in the particle-hole channel and a surface-volume, density-
dependent pairing energy density; see \cite{schunck_microscopic_2014} for 
details. The potential field $V(\qVec)$ is initially computed in a 
domain $\Omega_{0}$ characterized by
$(q_{20},q_{30}) \in [500; 60.10^3]\times [-92.10^3; +92.10^3]$ (in the units 
above). In practice, the HFB calculation did not converge for every point in the 
domain; the initial grid is thus irregularly spaced and contains $2 784$ fully converged points.
At each point, the collective inertia tensor $B^{(\qVec)}_{kl}$ is 
computed using the perturbative cranking approximation of the adiabatic 
time-dependent Hartree-Fock (ATDHF) theory \cite{baran2011}. We also determine 
at each point the expectation value $q_N$ of the Gaussian neck operator, which 
will be used to determine the scission hyper-surface.

The initial collective state $g(\qVec,0)$ is defined as the product of Gaussian 
functions in the $q_{20}$- and $q_{30}$-directions,
\begin{equation}
g(q_{20},q_{30},0) = 
\frac{1}{\sigma_{20}\sqrt{2\pi}}e^{\frac{1}{2}\left(\frac{q_{20} - q^{(\text{g.s.})}_{20}}{\sigma_{20}}\right)^2}
\times
\frac{1}{\sigma_{30}\sqrt{2\pi}}e^{\frac{1}{2}\left(\frac{q_{30} - q^{(\text{g.s.})}_{30}}{\sigma_{30}}\right)^2},
\end{equation}
where $\sigma_{20}=2 800$ fm$^{2}$ and $\sigma_{30}=6 000$ fm$^{3}$ are the 
widths of the Gaussian functions, and $q^{(\text{g.s.})}_{20} = 3 000$ fm$^{2}$, 
$q^{(\text{g.s.})}_{30} = 0$ fm$^{3}$ the coordinates of the ground-state. The 
advantage of such a choice is that the initial wave-packet is given 
analytically and does not yield additional numerical errors.

The resulting wave packet is then multiplied by a plane wave characterized by a 
wave vector $\boldsymbol{k}=(2.6.10^{-3}, 0)$. This last step gives the initial 
state an initial momentum toward positive elongations. It also boosts the 
average energy of the initial state up to roughly 500 keV above the first 
fission barrier.

In this work, the isoline $q_N=3.5$ mass units defines the scission hyper-surface 
in the collective space. The width of the absorption zone is set to $w=8.10^3$ 
in the system of units adopted. In this area, we impose an average absorption 
rate of $r=20.10^{22}\text{ s}^{-1}$. We solve Eq.(\ref{eq:evolution0}) up to a 
time $t_{max}=60.10^{-22}$ s. For the time step values used in table 
\ref{tab:errorYields} shown in section \ref{sec:Pu240Results}, 
this corresponds to 60000 and 120000 time iterations. At the end of the 
simulation, the probability for the system to populate post-scission 
configurations is approximately 30\%. We checked that stopping the time 
iterations after $t_{\text{max}}$ would not significantly change the fission 
fragments yields.

\subsection{Construction of the spatial mesh}
\label{sec:Pu240Mesh}

To minimize the computational cost, the time-evolution is not performed on a 
regularly meshed hypercube of the collective space. Instead, we use several 
techniques offered by the finite element method to generate a more efficient 
partition of the domain $\Omega$. We list below the various steps followed to 
produce the mesh:
\begin{enumerate}

 \item {\bf Delaunay triangulation - } We start with the initial rectangular, irregularly spaced  
 domain $\Omega_{0}$ mentioned in Sec. \ref{sec:Pu240Physics}. We generate a 
 first set of regularly spaced vertices in this domain with  
 $h \equiv h_{20}=3.1\times h_{30}$ the resolution of this new mesh $\Omega_{1}$. A Delaunay 
 triangulation provides a partition of  $\Omega_{1}$. For every simplex, we 
 choose an interpolation polynomial of degree one. The continuity of each field 
 is ensured by placing the three required nodes at the vertices. The values of the 
 input fields ($V(\qVec), B^{(\qVec)}_{kl}, \cdots$) are then evaluated at each 
 node by linear interpolation in $\Omega_0$.
 
 \item {\bf Absorption areas - } The mesh $\Omega_{1}$ is then extended with 
 60 new spatial steps in the lower $q_{20}$ values, 40 new spatial steps in the upper $q_{20}$ 
 values and 20 in both $q_{30}$ directions. These extensions define 
 the absorption band. By default, the input fields in the absorption band are 
 extrapolated as constants based on their values at the edges of the mesh 
 $\Omega_{1}$. The one exception is the potential field in the lower $q_{20}$ 
 region of the absorption band. In this case, we evaluate $V(\qVec)$ as a convex 
 parabola continuously connected at $q_{20}= 500$ fm$^{2}$. This simple 
 extrapolation prohibits the system to explore oblate shapes during the dynamics. 
 This extended mesh is denoted $\Omega_{2}$.
 
 \item {\bf h-refinement and coarsening - } We take advantage of the flexibility 
 of the finite element method to refine or coarsen the mesh $\Omega_{2}$ 
 depending on its relevance in the time evolution of the system and the flux 
 calculation. This step is needed to improve the numerical precision of the 
 calculation while  keeping the computational cost as low as possible. After 
 this series of refinement, the new mesh is $\Omega_{3}$.
 
 \begin{itemize}
 
   \item In the regions of the domain $\Omega_{2}$ where the potential takes 
    very large values, the collective wave-function $g(\qVec,t)$ will remain very 
    small during the entire time evolution. We use the criteria $V>V_{GS}+35$ 
    Mev, $q_{N} < 1.0$ and $A_{H} > 170$ to automatically detect such regions, 
    where we locally coarsen the mesh by discarding two out of three vertices.
  
    \item By contrast, other regions of the domain $\Omega_{1}$ are important to 
    the physics of fission, such as near the ground-state (where the initial 
    collective state is defined) or near the saddle points. In these regions, we 
    apply an additional h-refinement step: each simplex is divided in four new 
    elements by adding a vertex in the middle of each of its edges. At the 
    boundaries of this refinement zone, some simplices may be divided differently 
    to ensure the continuity of the fields in the new mesh.
 
    \item The most critical region of the potential energy surface is the 
    scission hyper-surface, where one calculates the total flux. In this area, 
    it is essential to maximize the accuracy of the calculation of the function 
    $g(\qVec,t)$ and its derivatives. Therefore, we apply two successive h-
    refinement steps near the scission hyper-surface.
 
    \item Up to now the domain $\Omega_{2}$ remains rectangular. However, the 
    areas past and far away the scission hyper-surface are totally irrelevant to 
    the calculation. We thus crop the mesh to retain only the regions of 
    interest. This is done with the following criterion: a simplex in the 
    external region is kept only if its distance to the scission hyper-surface is 
    lower than 10$^3$.
    
 \end{itemize}
 
 \item {\bf p-refinement - } We apply one p-refinement step to the whole mesh 
 $\Omega_{3}$, that is, after local h-refinement and careening and cropping. 
 As a result, the degree of interpolation polynomials in each simplex increases 
 to two. This requires three new nodes per simplex, which are positioned in the 
 middle of the simplex edges. The resulting, and final mesh, is denoted $\Omega$. 
 Starting with $h=397$ (in fm units), it contains a list $L$ of about $106 000$ different nodes
 
 \item {\bf Evaluation of fields - } Finally, the input fields are again 
 evaluated at each node of $\Omega$. To do so, results from the original  
 calculation on the mesh $\Omega_{0}$ are linearly interpolated as discussed in 
 the first step.
 
\end{enumerate}

\begin{figure}[!ht]
  \begin{center}
  \includegraphics[width=0.45\textwidth]{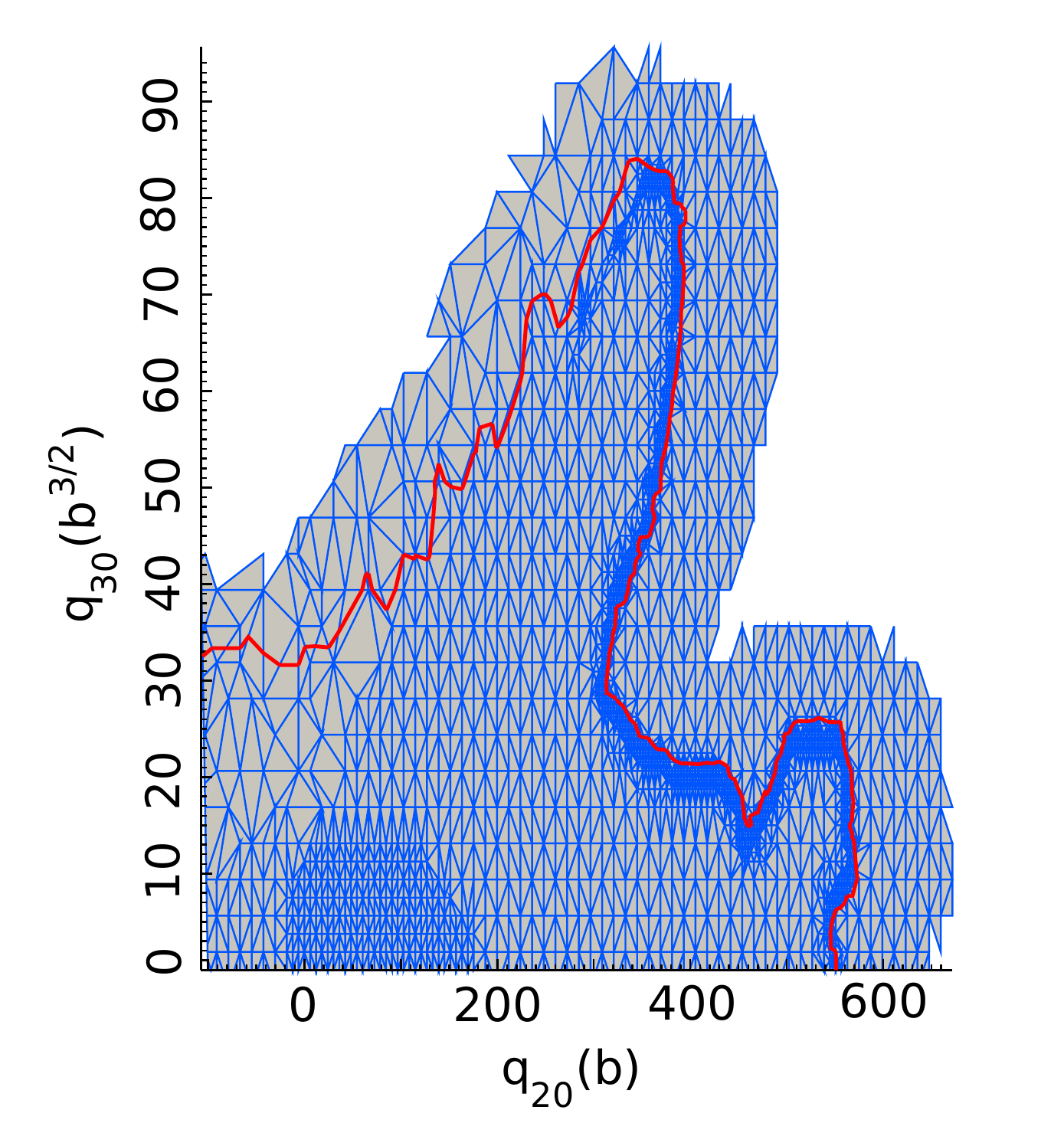}
  \caption{Optimized mesh $\Omega$ obtained with a parameter $h=1190$. The red 
  line represents the  scission hyper-surface, defined arbitrarily by the isoline 
  $q_N=3.5$. This figure shows only the region $q_{30}>0$ fm$^{3}$ and $q_{20}>-10^{5}$ 
  fm$^{2}$.}
  \label{fig:mesh}
  \end{center}
\end{figure}

The figure~\ref{fig:mesh} shows the final mesh $\Omega$ obtained after all the 
aforementioned steps. The main advantage of mesh optimization is to reduce 
significantly the total number of nodes required to achieve a given numerical 
precision. As an example, let us consider a two-dimensional, 
regularly-spaced hypercube mesh $\mathcal{M}_{476}$ with $h=476$ and degree 2 polynomials.
We compare it with an optimized version $\Omega$ built from the same initial spatial step. The total number of nodes has been 
decreased by more than 25\% even though the mesh is locally up to four times more refined in the 
physically relevant areas.

\subsection{Results}
\label{sec:Pu240Results}

We performed a series of calculations with different values of the spatial and 
time resolution parameters $h$ and $\delta t$. The parameter values $h = 793, 
595, 476, 397, 340, 298, 264 $ allow us to analyze the convergence in space, 
while the different time-steps $\delta t= 10^{-3}, 5.10^{-4}$  (in $10^{-22}$s) 
control the convergence in time.

\begin{figure}[!ht]
  \begin{center}
  \includegraphics[width=0.4\textwidth]{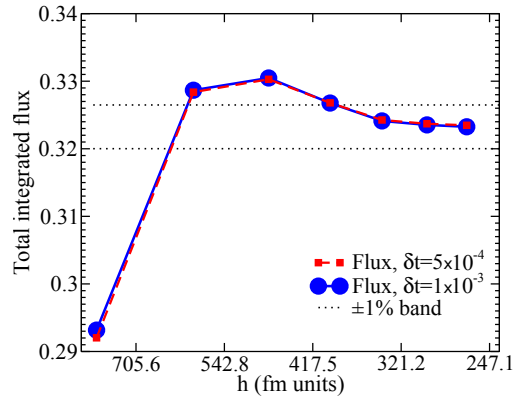}
  \caption{Total flux crossing the scission hyper-surface in the interval 
  $[0; t_{max}]$ for two different time-steps $\delta t= 10^{-3}, 5.10^{-4}$ (in 
  $10^{-22}$s). The flux is computed for several meshes characterized by the 
  parameter $h = 793, 595, 476, 397, 340, 298, 264 $. Note that the x-axis is 
  in log scale.}
  \label{fig:fluxTot}
  \end{center}
\end{figure}

We first consider the total cumulative flux that crosses the scission 
hyper-surface during the whole evolution time. This total flux $F(t_{max})$ 
reads
\begin{equation}
 F(t_{max})= \sum_\xi F(\xi,t_{max})
\end{equation}
The figure~\ref{fig:fluxTot} shows the rate of convergence of the flux as a 
function of spatial resolution. The relative difference between the values obtained for $h=298$ and $264$ is less than $10^{-3}$. We can also notice that 
the calculation is fully converged in time: the error on the flux is mostly 
driven by the spatial resolution.

Last but not least, we computed the fission fragment mass yields as a function 
of both time and space resolutions. As discussed in \cite{younes_fragment_2012}, 
the mass of the fission fragments along the scission hyper-surface is not 
necessarily an integer, since both the compound nucleus and the fission fragment 
are described in the Hartree-Fock-Bogoliubov approximation, where particle 
number is not conserved. As a result, one must take into account the fluctuations 
in particle numbers when comparing the yields with experimental data. Following 
\cite{younes_fragment_2012}, we have convoluted the yields coming out of the 
flux calculation with a Gaussian with a width of $\sigma = 3.5$ mass units. As 
customary, the resulting values are then normalized to 200. To measure the 
convergence of the yields, we define the quantity $e_{\text{Y}}$
\begin{equation}
e_{\text{Y}} = || Y(A) - Y_{\text{ref}}(A) ||_{\infty}.
\end{equation}
The most accurate calculation was obtained with $h=264$ and $\delta t=5.10^{-4}$ 
and is chosen as the reference $Y_{\text{ref}}(A)$.

\begin{table}[!ht]
\begin{center}
$
 \begin{array}{c|ccccccc} 
 \hline

      \delta t\,  | \,h   &  793 & 595 & 476 & 397 & 340 & 298 & 264 \\
 \hline
 10^{-3}     &  1.10 & 0.69 & 0.28 & 0.26 & 0.13 & 0.07 & 0.03 \\
 5.10^{-4}   &  1.08 & 0.67 & 0.27 & 0.24 & 0.11 & 0.05 & 0.00 \\
 \hline
 \end{array}
 $
 \caption{Deviation $e_{\text{Y}}$ as a function of time and space resolutions. 
 The parameters $\delta t$ and $h$ are expressed in $10^{-22}s$ and Fermi units respectively.}
 \label{tab:errorYields}
 \end{center}
\end{table}

We compute the deviation $e_{\text{Y}}$ for the different values of our 
numerical parameters. The results are summarized in table~\ref{tab:errorYields}.
We first note that the differences on the deviation $e_{\text{Y}}$ caused by 
the time resolution are of the order of 0.02. They are much smaller than the 
variations induced by the spatial resolution. The evolution of $e_{\text{Y}}$ as 
a function of $h$ shows the convergence in space of our calculation. The values 
obtained for $h=298$ typically correspond to a numerical error of a few percents for masses with $Y(A)>1\%$.

\section{Program \theCode}
\label{sec:programmTheCode}

The package \theCode \ is composed of the following directories and files:
\begin{itemize}
 \item {\tt README}: contains detailed instructions to build the solver, the tools and their dependencies, and to run the code with the examples provided; 
 \item {\tt Makefile}: a standard GNU makefile to build the solver and the tools;
 \item {\tt src/}: C++ source files of the TDGCM solver and of the tools;
 \item {\tt tools/}: additional C++ source files, python and shell scripts to handle the inputs and outputs of the TDGCM solver;
 \item {\tt benchmarks/}: a few preset inputs and their corresponding outputs;
 \item {\tt doc/}: documentation of the package in DoxyGen format.
\end{itemize}

The full Felix package depends on several standard Open Source libraries:
\begin{itemize}
 \item The TDGCM solver itself requires BLAS, LAPACK, and a Fortran compiler with 
OpenMP support;
 \item The documentation requires DoxyGen-1.8.6 or higher;
 \item In order to build the full set of tools included in this release, the user must also install GSL, PETSc, SLEPc and Boost. The versions GSL-0.16, PETSc-3.5.2, SLEPc-3.5.3 and Boost-1.54 have been used during development.
\end{itemize}

\subsection{Compilation}

The program is shipped with a Makefile containing a preset configuration assuming  compilation with the GNU gcc compiler on a standard LINUX distribution. The user should expect to have to change this Makefile to match his/her own system configuration. The Makefile contains some instructions to help the user with this configuration step. The different components of the package can be compiled separately by typing the following commands:
\begin{itemize}
 \item {\tt make Doc}: generate the DoxyGen documentation in the directory, which is located in {\tt doc/DoxygenDoc}.
 \item {\tt make Solver}: build the TDGCM+GOA solver executable. Its name is {\tt tdgcmFE} and it is located by default in {\tt src}.
 \item {\tt make Tests}: build the executable for the full suite of tests included in the package. The name of the executable is {\tt tdgcmFEtest} and is located in {\tt tests/src/}.
 \item {\tt make Tools}: compile all tools in the directory {\tt tools/}.
\end{itemize}

\subsection{Running the solver}

There are two different ways to run the \theCode \ solver with specific input data. 
\begin{itemize}
\item If an option file {\tt input.opt} is available, the user may simply type

{\tt ./src/tdgcmFE  input.opt}

\item Otherwise, the list of options can directly be passed via the command line as:

{\tt ./src/tdgcmFE  \ -\,-option0 [value0] \ -\,-option1 [value1] ...}

\end{itemize}
The available options are discussed below in section \ref{subsec:options}.

By default, the solver uses every available core on the machine. As usual, the number of OpenMP threads can be controlled by setting the environment variable {\tt OMP\_NUM\_THREADS}. 

\subsection{System of units}
\label{subsec:unitSystem}

By default, the value of the reduced Planck constant $\hbar$ is set to
\begin{equation} 
\hbar= 6.58211928 \quad (10^{-22}\text{MeV.s}).
\end{equation}
This imposes a consistency relation between the energy and time units that can be used in the code. The most natural choice is to set the value of energies in MeV so that time is given in units of $10^{-22}$ s. This is recommended and is the default setting for \theCode. 

If needed, the value of the reduced Planck constant can be changed in the file {\tt./src/defines.h}, which allows the user to define his/her own set of physical units. Note that the code must be entirely re-compiled for such changes to take effect. Also, special attention must be paid to setting the units for the inertia tensor and coordinates $\qVec$ in a consistent manner.

\section{Inputs and outputs}
\label{sec:inputsOutputs}

\subsection{Input files}
\label{subsec:inputFile}

\theCode\ reads most of its input from several ASCII files. The names of these files begin with the same user-defined prefix, and have a specific extension, as for instance, 
{\tt example.\-coor}, {\tt example.\-val}, {\tt example.\-geo}, etc. The following mandatory files are needed by the solver:
\begin{itemize}
\item {\tt example.coor}: This file contains an unsorted list $L$ of points in the domain $\Omega$ defined by their coordinates $\qVec$ (in their appropriate unit). In the other inputs files, a point is labeled by its index in the list $L$. Each line of this file contains a series of $N$ {\tt double} numbers separated by a space;
\item {\tt example.geo}: This file defines the list of all simplices of the mesh. Every line contains a series of {\tt int} integers separated by spaces: the first integer is the degree of the interpolating polynomial used in the simplex; it is followed by the list of the $N+1$ vertices of each simplex, followed by all the nodes used to define the interpolating polynomials.
In this version of \theCode \, the only limitation on the geometry comes from the boundary condition Eq.~\ref{eq:dirichlet}.
Internally, this condition is imposed by setting the values of the coefficients $g(\qVec_j,t)$ to zero for every node lying on the boundary $\partial \Omega$.
The user must therefore design the boundary elements so that the aforementioned property implies Eq.~\ref{eq:dirichlet}.
\item {\tt example.val}: This file contains the values of several fields at each point of the list $L$. The first line starts with the special character {\tt \#}, and contains a list of {\tt string} keys that define the fields. Mandatory fields are the potential $V(\qVec)$ and the lower part of the inertia tensor. The key for the potential is simply {\tt v}; for the inertia tensor: in 2D, the keys are {\tt B00}, {\tt B10}, {\tt B11}; in 3D, they are {\tt B00}, {\tt B10}, {\tt B20}, {\tt B21}, {\tt B22}; etc. Optional fields recognized by the code  are {\tt qN} -- the expectation value of the Gaussian neck operator, and {\tt AH} -- the mass of the heaviest fragment. 
\item {\tt example.init}: It contains the values of the initial wave function $g(\qVec, t=0)$ at each point of the list $L$. Each line is made of two {\tt double} corresponding to the real and imaginary parts of the wave function.
\end{itemize}

The user may also specify additional features and options. They are handled through the option list described in section~\ref{subsec:options} below. Some of these options require one or several additional input files:
\begin{itemize}
\item {\tt example.front}: This file contains a list of oriented hyper-surfaces for which the solver will calculate the flux. The surfaces are simplex edges and are defined by $N$ vertices. The additional vertex "vOpposite" shown in figure~\ref{fig:vOpposite} must be specified to set up the orientation.
\item {\tt example.matM} and {\tt example.matH}: These two files store the sparse matrices $M$ and $H$ defined in Eq.~(\ref{eq:matrixElements}).
\item {\tt example.opt}: A file containing the list of options described in section~\ref{subsec:options}.
\end{itemize}

\subsection{Options}
\label{subsec:options}

The user can pass a number of options to the solver. Most of these options control the numerical parameters of the calculation and the frequency at which output data is written on disk.\\

{\noindent\tt --help} (flag) \\ 
If present, the code only prints a help message and stops the execution.
Default: Absent.\\

{\noindent\tt --version} (flag) \\ 
If present, the code only prints the version number and stops the execution.
Default: Absent.\\

{\noindent\tt --file} ({\tt string}) \\ 
Prefix name for the input files. 
All input files must be named as {\tt prefix.ext} where {\tt ext} is one of the extensions described in section~\ref{subsec:inputFile}.
Default: {\tt input}.\\

{\noindent\tt --outputDir} ({\tt string}) \\ 
Name of the output directory.
If no existing directory is found, a new directory is created.
Default: {\tt./results}\\

{\noindent\tt --dump} ({\tt integer}) \\ 
Number of time iterations between two prints of the solution and the flux.
Default: 100.\\

{\noindent\tt --refresh} ({\tt integer}) \\ 
Number of time iterations between two displayed lines in the standard output.
Default: 100.\\

{\noindent\tt --max} ({\tt integer}) \\ 
Maximum number of time iterations for the calculation.
Default: -1 (no maximum).\\

{\noindent\tt --step} ({\tt double}) \\ 
Time step of the calculation. By default, the unit of the time step is 10$^{-22}$ s.
Default: $10^{-4}$.\\

{\noindent\tt --inversionTol} ({\tt double}) \\ 
Numerical tolerance for matrix inversion; see Eq.~(\ref{eq:convergenceCriteria}).
Default: $10^{-15}$.\\

{\noindent\tt --limit} ({\tt integer}) \\ 
Maximum number of iterations in matrix inversions.
Default: $10^{5}$.\\

{\noindent\tt --absRate} ({\tt double}) \\ 
Average absorption rate per time unit in the absorption zone. For instance, 
{\noindent\tt --absRate 20} specifies an absorption rate of 20.10$^{22}$ 
s$^{-1}$, since the basis time unit is 10$^{-22}$ s
Default: 0.\\

{\noindent\tt --absWidth} ({\tt double}) \\ 
Width of the absorption zone. The unit depends on the units of the collective coordinates $\qVec$. For any node of the mesh, the euclidean distance $d$ to the boundary $\partial \Omega$ is computed. A node is included in the absorption zone if and only if $d<${\tt absWidth}. A negative value will lead to no absorption. Example: Consider a rectangular 1D space with the axial quadrupole moment given in b. Assume the domain is $q_{20} \in [0, 600]$ b and {\tt absWidth}=10. Then, all points with $q_{20}\in[ 0, 10] \cup [ 590, 600] $ b will be included in the absorption band. 
Default: -1.\\

{\noindent\tt --calcEnergy} (flag) \\ 
If this flag is present, the code prints the average energy of the solution every {\tt dump} time iteration.
Default: Absent.\\

{\noindent\tt --dumpMat} (flag) \\ 
If present, the sparse real matrices $M$ and $H$ are stored in the files {\tt input.matM} and {\tt input.matH} respectively.
Default: Absent.\\

{\noindent\tt --readMat} (flag) \\ 
If present, the code reads the $M$ and $H$ matrices from the files {\tt input.matM} 
and {\tt input.matH}. Note that these matrices depend only on the particular mesh of 
the collective space, but not on time $t$: for a given mesh, they can be pre-calculated, 
stored on disk using the {\noindent\tt --dumpMat} flag, and re-used in a different 
run. 
Default: Absent.\\

{\noindent\tt --ISMethod} ({\tt string}={\tt 'file'}, {\tt 'impulsed'} or {\tt 'wavePacket'}) \\ 
Method of determination of the initial state $g(\qVec, t=0)$. 
\begin{itemize}
\item If set to {\tt 'file'}, the solver reads the initial state from the file {\tt input.init}.
\item If set to {\tt 'impulsed'}, the code reads the initial state contained in the file {\tt input.init} and multiplies it by a plane wave.
We note ${\bf k} = k \hat{\bf{k}}$ the wave vector characterizing the plane wave, with $\hat{\bf{k}}$ the unit vector indicating its direction and $k$ the modulus. 
The user must specify the coordinates of $\bf{k}$ in a file {\tt input.k}.
The modulus $k$ is computed by the solver so that the average energy of the initial state matches the value provided with the option {\tt --ISEnergy}.
The user may provide an initial guess of the parameter $k$ via the option {\tt --ISLambdaGuess}.
\item If set to {\tt 'wavePacket'}, the solver builds a linear combination of states provided by the user $g(\qVec, t=0)=\sum_k \alpha_k g_k(\qVec)$. 
The user provides the states $g_k(\qVec)$ as a set of files {\tt state\_0}, {\tt state\_1}, etc. 
A single directory, the name of which is set with the option {\tt ISStatesDir}, must contain all the files.
The format for the files {\tt state\_k} is the same as for {\tt input.init}.
In the expansion of the initial state, all weights have a Gaussian dependency on the expectation value $E_k$ of the energy of each state: $\alpha_k \propto \operatorname{exp}(-(E_k -\langle E \rangle)^2/2\sigma^2)$.
The user can tune the parameter $\sigma$ via the option {\tt ISSigma}.
The code determines the parameter $\langle E \rangle$ so that the energy of the wave packet matches the value provided in the option {\tt --ISEnergy}.
\end{itemize}
Default: {\tt 'file'}.\\

{\noindent\tt --ISEnergy} ({\tt double}) \\ 
Requested energy of the initial state for the methods {\tt 'impulsed'} and {\tt 'wavePacket'}.
Default: 0.\\

{\noindent\tt --ISSigma} ({\tt double}) \\ 
Gaussian width of the initial state for the method {\tt 'wavePacket'}.
Default: 1.\\

{\noindent\tt --ISStatesDir} ({\tt string}) \\ 
Directory containing the files {\tt state\_k} required for the cconstruction of the initial state with the method {\tt 'wavePacket'}.
Default: {\tt './eigenStates'}.\\

{\noindent\tt --ISLambdaGuess} ({\tt double}) \\ 
Initial guess for the modulus $k$ of the plane wave multiplying the initial state with the method {\tt 'impulsed'}.
If this value is negative, the code will initialize $k$ with a default value.
Default: -1.\\

{\noindent\tt --ISNorm} (flag) \\ 
If present, the initial state will be normalized to 1 before starting time iterations.
Default: Absent.\\

{\noindent\tt --frontier} ({\tt string}={\tt 'qN'} or {\tt 'file'}) \\ 
Method of determination of the frontier. 
If set to {\tt 'file'}, the frontier will be read from the corresponding input file. 
If no such file can be found, the frontier is empty and the instantaneous flux is zero. 
If set to {\tt 'qN'}, the frontier will be computed on the fly as an isoline of the field {\tt 'qN'}. 
This field must then be present as an additional column in the {\tt .val} file, with the key {\tt qN}.\\
Default:{\tt 'file'}.\\

{\noindent\tt --frontIso} ({\tt double}) \\ 
Value of the field {\tt qN} used to define the scission hyper-surface.
Default:1.\\

{\noindent\tt --fluxInst} (flag) \\ 
If present, the instantaneous flux through the frontier is recorded every {\tt dump} time  iteration.
Default: Absent.\\

{\noindent\tt --lumpedMass} (flag) \\ 
If present, the lumped mass approximation is applied when calculating the $M$ matrix.
Default: Absent.\\

{\noindent\tt --pCoeff} ({\tt double}) \\ 
Arbitrary multiplicative factor applied to the potential field $V(\qVec)$.
Default:1.\\

{\noindent\tt --bCoeff} ({\tt double}) \\ 
Arbitrary multiplicative factor applied to the inertia tensor field $B_{kl}^{(\qVec)}$.
Default:1.

\subsection{Output files}

All outputs are recorded in a directory that can be specified via the option {\tt outputDir} described in the previous section. Upon successful execution of the solver, this directory should contain
\begin{itemize}
\item The file {\tt example.opt}: It contains the list of all options used for the run. It could be re-\/used as an input file for any other calculation;
\item The directory {\tt gFunction/}: Each file {\tt g\-Function.xxx.log} in 
this directory contains the solution $g$ after xxx time iterations. These 
files are formatted in the same way as the input file {\tt example.init}. 
For example, the file {\tt gFunction.000000000.log} is a copy of 
{\tt example.init}, unless a renormalization has been requested by the user 
through the use of option {\tt norm} described above. Thanks to this identical 
format, these files can be used in subsequent runs as initial wave-functions 
by simply copying them in place of whatever {\tt .init} file was used. The 
program {\theCode} thus has basic checkpointing capabilities;
\item The file {\tt normDeviation}: It contains the deviation $||g(t)||_2 - ||g(t=0)||_2$ as a function of the number of time iterations;
\item The file {\tt example.front}: This file contains the frontier used for the calculation. It could be either a copy of the corresponding input file or the result produced by the solver itself if the frontier is defined from a {\tt qN} isoline with the option {\tt --frontIso='qN'}.
\item The directory {\tt flux/}: Each file {\tt flux.\-xxx.\-log} in this directory contains the time integrated flux $F(\xi)$ on each element $\xi$ of the frontier after xxx time iterations. 
\end{itemize}

Some additional outputs may appear depending on the options provided:
\begin{itemize}
\item A file {\tt averageEnergy.log}: This file contains the average energy of the solution as a function of the number of time iterations;
\item A file {\tt frontNorm.log}: It contains the coordinates of the normal vector to each hyper-surface of the frontier;
\item A series of files {\tt fluxInst.XXX.log} in the {\tt flux/} directory: These files contain the instantaneous flux at the frontier. This flux is not integrated over time.
\end{itemize}

\section{Acknowledgements}
\label{sec:acknowledgement}

The research was carried out under the US-France International Agreement on Cooperation on Fundamental Research Supporting Stockpile Stewardship.
This work was partly performed under the auspices of the U.S.\ Department of Energy by Lawrence Livermore National Laboratory under Contract DE-AC52-07NA27344. Computational resources were provided through an INCITE award ``Computational Nuclear Structure'' by the National Center for Computational Sciences (NCCS) and National Institute for Computational Sciences (NICS) at Oak Ridge National Laboratory, and through an award by the Livermore Computing Resource Center at Lawrence Livermore National Laboratory.





\section{References}
\bibliographystyle{elsarticle-num}
\bibliography{biblio}







\end{document}